\documentclass{article}
\usepackage{amssymb}
\usepackage{multirow}
\usepackage{subfig}
\usepackage{float}
\usepackage{booktabs}
\usepackage{hyperref}
\hypersetup{
hidelinks,
colorlinks=true,
linkcolor=red,
citecolor=cyan,
urlcolor = magenta
}
\usepackage{spconf,amsmath,graphicx}

\title{LVC-LGMC: Joint Local and Global Motion Compensation \\
for Learned Video Compression}
\name{Wei Jiang, Junru Li, Kai Zhang, Li Zhang}
\address{Multimedia Lab, Bytedance Inc., San Diego CA. 92122 USA\\\tt \normalsize \url{wei.jiang1999@outlook.com}}
\begin{document}
%
\maketitle
\begin{abstract}
Existing learned video compression models employ flow net or 
deformable convolutional networks (DCN) to estimate motion information.
However, the limited receptive fields of flow net and DCN inherently direct their attentiveness towards the \textit{local} contexts.
\textit{Global} contexts, such as large-scale motions and global correlations among frames are ignored, presenting
a significant bottleneck for capturing accurate motions. 
To address this issue, 
we propose a joint \textit{local} and \textit{global} motion compensation module (LGMC) for leaned video coding. 
More specifically,
we adopt flow net for \textit{local} motion compensation.
To capture \textit{global} context, we employ the cross attention in feature domain for motion compensation. 
In addition, to avoid the quadratic complexity of vanilla cross attention,
we divide
the softmax operations in attention into two independent softmax operations, leading 
to linear complexity. To validate the effectiveness of our proposed LGMC, 
we integrate it with DCVC-TCM and obtain learned video compression with joint 
\textit{local} and \textit{global} motion compensation (LVC-LGMC).
Extensive experiments demonstrate that our LVC-LGMC has significant rate-distortion performance improvements over baseline DCVC-TCM.

\end{abstract}
\begin{keywords}
Motion estimation, neural video compression
\end{keywords}
\section{Introduction}
The rapid development of social media and video applications triggers the increases of the video date volume, bringing challenges to video compression~\cite{lu2019dvc,li2021deep}.
In recent years, learned video compression has attracted lots of attentions. 
Most learned video compression models~\cite{lu2019dvc,li2021deep,lin2022dmvc,guo2023enhanced} are based on predictive coding
paradigm, which employs a flow net or Deformable Convolutional Networks (DCN) ~\cite{dai2017deformable} to predict the motion information between 
the reference frame and current frame. Subsequently, a motion codec is cooperated to 
compress the motion information. Meanwhile, residual codec~\cite{lu2019dvc} or a 
contextual codec~\cite{li2021deep} is used to compress the residuals or contexts.
To enhance the compression performance, advanced entropy models~\cite{li2022hybrid} and flow coding techniques~\cite{hu2022coarse} are investigated in recent years. Some models~\cite{li2023neural} even outperform the Versatile Video Coding (VVC) under the Low Delay configuration.

Traditional video coding is block-based, which is capable of search over a predefined region for motion estimation.
Flow-based or DCN-based motion estimation can only 
handle small motions due to the limited receptive
field of Convolution Neural Networks (CNN), which can only capture
\textit{local} redundancy.
\textit{Global} redundancies are existed even in the scenario of small motions, which has not been sufficiently explored in learned video compression.

To address above issues, we propose a novel mixed flow-attention based 
motion estimation and motion compensation module in feature space
for joint \textit{local} and \textit{global} redundancy capturing.
Our module is built upon the conditional coding paradigm~\cite{li2021deep,sheng2022temporal}.
We integrate the proposed joint \textit{local} and \textit{global} module to DCVC-TCM
to obtain the learned video compression model LVC-LGMC.
In the LVC-LGMC, multi-scale motion compensation is adopted.
More specifically, when compressing the current P-frame $\boldsymbol{x}_t$, we use a flow net to estimate the motion offset. 
By warping the multi-scale features using the decoded estimated offset, the local redundancies between frames can be well captured.
To achieve global motion 
estimation,
we propose to employ the cross attention between the propagated features 
and middle features of current frame for the long-range modeling.
The attention map contains the global 
similarity between frames without consuming additional bits for representation.
However, the quadratic complexity of vanilla attention impedes the compression of high-resolution videos.
To address this issue, we propose to divide the softmax in 
vanilla attention into two softmax operations~\cite{shen2021efficient}.
The dot product features after two independent softmax operations is treated as the similarity metric. Such division of softmax operation makes the complexity increases
linearly with the resolution, leading the operation more efficient.
When compared with the baseline DCVC-TCM~\cite{sheng2022temporal}, the
proposed LVC-LGMC reduces $10$\% bit-rates on MCL-JCV test sequences~\cite{wang2016mcl}.\par
The contributions of this paper are summarized as follows:
\begin{itemize}
    \item We propose a novel attention-based motion compensation module 
    to handle large-scale movements and capture global redundancy between frames but with linear complexity and without extra bits.
    To our knowledge, this is the first attempt to use cross attention
    for motion compensation.
    \item We incorporate proposed attention-based motion compensation with 
    flow-based motion compensation for joint local and global motion compensation.
    The proposed method significantly boosts the model performance.
    \end{itemize}

\section{The Proposed LVC-LGMC Method}
\label{sec:method}
\begin{figure}
    \centering
    \includegraphics[width=\linewidth]
    {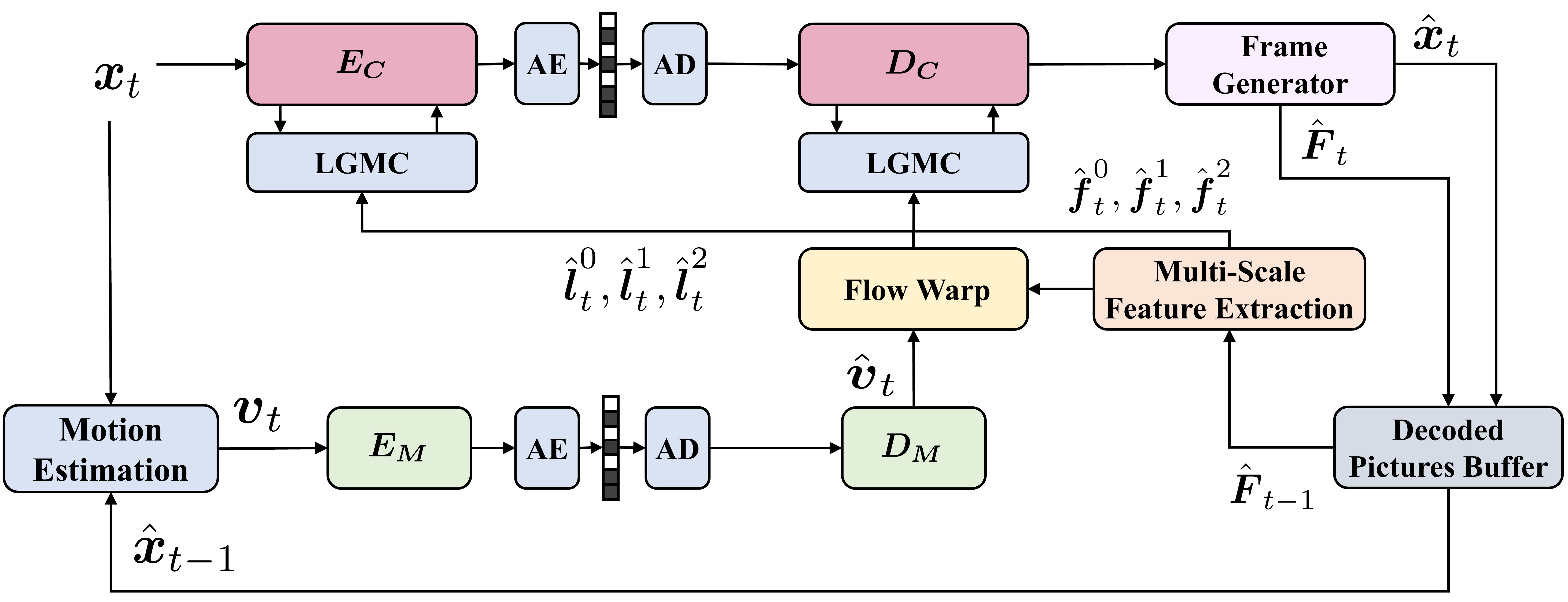}
    \caption{Overall frame work of the proposed LVC-LGMC. $\boldsymbol{E}_C$ is the contextual encoder, and $\boldsymbol{D}_C$ is the contextual decoder. $\boldsymbol{E}_M$ is 
    the MV encoder, and $\boldsymbol{D}_M$ is the MV decoder. LGMC is the proposed joint local and global motion compensation module.}
    \label{fig:arch}
\end{figure}

The paradigm of the proposed LVC-LGMC is illustrated in Fig.~\ref{fig:arch}.
We reproduce the DCVC-TCM~\cite{sheng2022temporal} and build LVC-LGMC on top of it.
Following DCVC-TCM,
we adopt temporal propagated multi-scale features for local compensation.
\subsection{Flow-based Local Compensation}

When compressing the $t$-th frame $\boldsymbol{x}_t$, first, we conduct the local compensation~\cite{sheng2022temporal}. In particular, 
multi-scale features $\hat{\boldsymbol{f}}_t^0, \hat{\boldsymbol{f}}_t^1,\hat{\boldsymbol{f}}_t^2$ are extracted 
from propagated feature $\hat{\boldsymbol{F}}_{t-1}$.
Motion vector $\hat{\boldsymbol{v}}_t$ is employed to warp the multi-scale features
to multi-scale local contexts $\hat{\boldsymbol{l}}_{t}^0,\hat{\boldsymbol{l}}_{t}^1,\hat{\boldsymbol{l}}_{t}^2$.
The $\hat{\boldsymbol{l}}_{t}^0,\hat{\boldsymbol{l}}_{t}^1,\hat{\boldsymbol{l}}_{t}^2$
are concatenated with the current frame $\boldsymbol{x}_{t}$, middle-feature $\boldsymbol{y}^1_t$
and middle-feature $\hat{\boldsymbol{y}}_{t}^2$ respectively.
As such, the network could well understand 
how to conduct conditional coding.
During decoding, the multi-scale contexts are also concatenated to recover the frame.
The overall process can be formulated as:
\begin{equation}
    \hat{\boldsymbol{x}}_{t} = D_C\left(\left\lfloor E_C(\boldsymbol{x}_t|\hat{\boldsymbol{l}}_{t}^0,\hat{\boldsymbol{l}}_{t}^1,\hat{\boldsymbol{l}}_{t}^2)\right\rceil|\hat{\boldsymbol{l}}_{t}^0,\hat{\boldsymbol{l}}_{t}^1,\hat{\boldsymbol{l}}_{t}^2\right),
\end{equation}
where $E_C$ and $D_C$ are the contextual encoder and the contextual decoder.

\subsection{Attention-based Global Compensation}
\begin{figure}
    \centering
    \includegraphics[width=\linewidth]
    {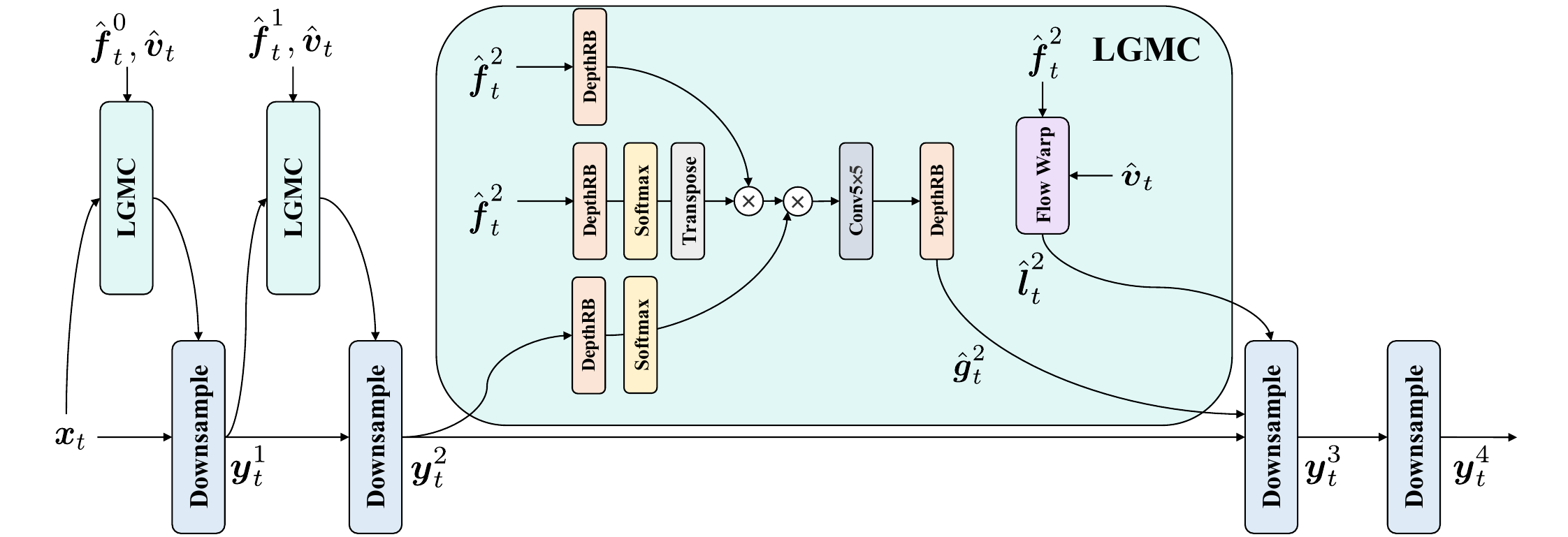}
    \caption{Illustration of the joint local and global motion compensation module (LGMC) at encoder side.}
    \label{fig:lmc}
\end{figure}
\begin{figure}
    \centering
    \includegraphics[width=\linewidth]
    {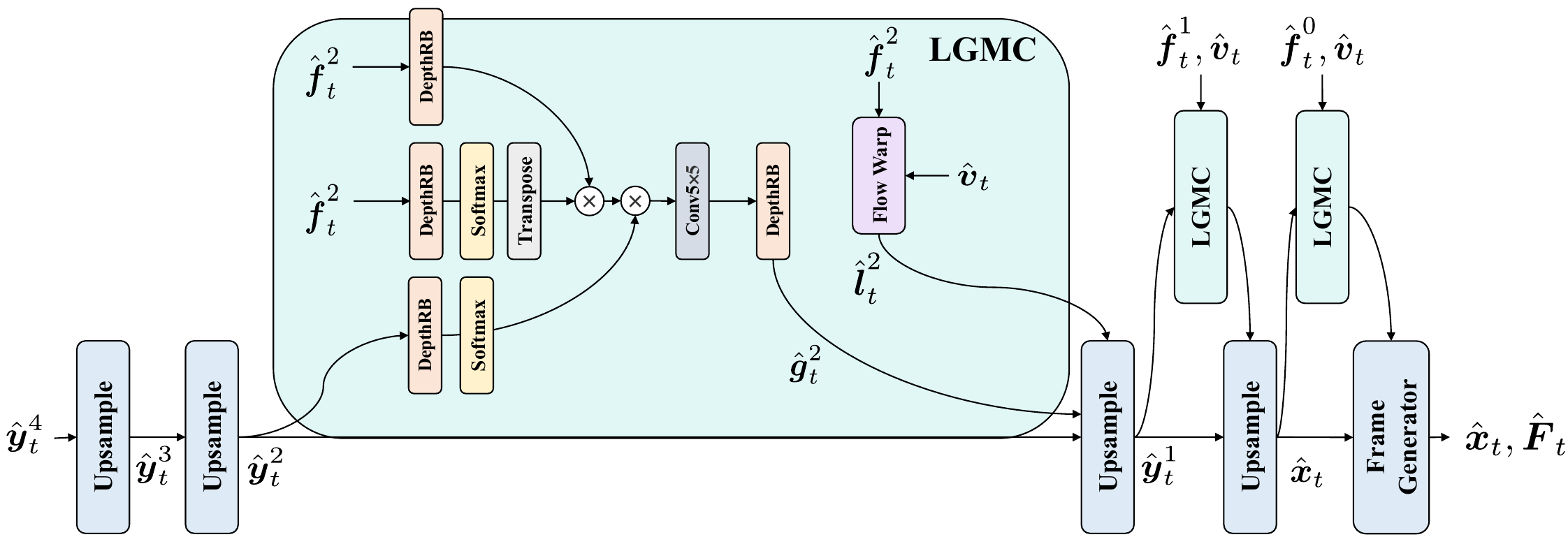}
    \caption{Illustration of the joint local and global motion compensation module (LGMC) at decoder side.}
    \label{fig:lmc_dec}
\end{figure}
The \textit{vanilla} approach for global compensation is introduced first.
Due to the limitations of flow-based motion compensation, 
global motion compensation is utilized for learned video compression.
It has been recognized that the CNN has limited receptive fields, such that
global attention is adopted for global motion compensation.
Regarding the compression of $\boldsymbol{x}_t$,
given multi-scales features $\hat{\boldsymbol{f}}_t^0, \hat{\boldsymbol{f}}_t^1,\hat{\boldsymbol{f}}_t^2$,
 current frame $\boldsymbol{x}_t$, and middle features $\boldsymbol{y}_t^1, \boldsymbol{y}_t^2$, we take $\boldsymbol{y}_t^2 \in \mathbb{R}^{C\times \frac{L}{16}}, \hat{\boldsymbol{f}}_t^1 \in \mathbb{R}^{C\times L}$ as an example.
where $C$ is the channel number, and $L=H\cdot W$, $H$ is the height and $W$ is the width of the frame.
The $\boldsymbol{y}_t^2$ and $\hat{\boldsymbol{f}}_t^2$ are first fed into a Depth Residual Bottleneck~\cite{jiang2023slic} for nonlinear embedding.
The \textit{vanilla} approach adopts cross attention~\cite{vaswani2017attention,jiang2022mlic},
which is formulated as:
\begin{equation}\label{eq:vanilla}
    \hat{\boldsymbol{g}}_{t}^2 = \underbrace{\textrm{softmax}\left(\boldsymbol{y}_t^2 \left(\hat{\boldsymbol{f}}_t^2\right)^\top\right)}_{\textrm{non-negative}}\hat{\boldsymbol{f}}_t^2.
\end{equation}
\begin{figure*}[t]
    \centering
    \subfloat{
      \includegraphics[scale=0.35]{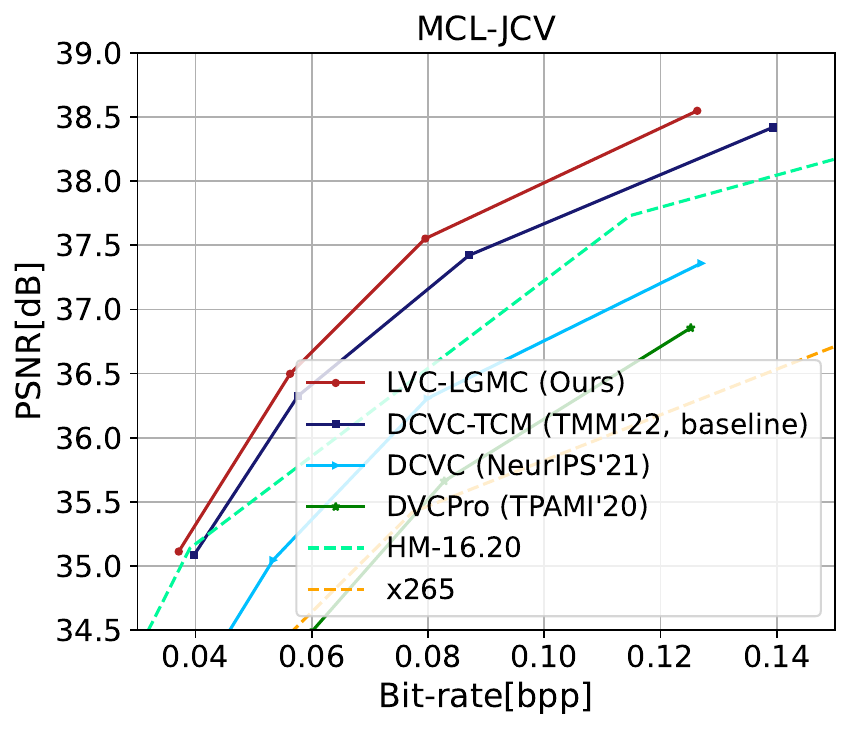}}
    \subfloat{
      \includegraphics[scale=0.35]{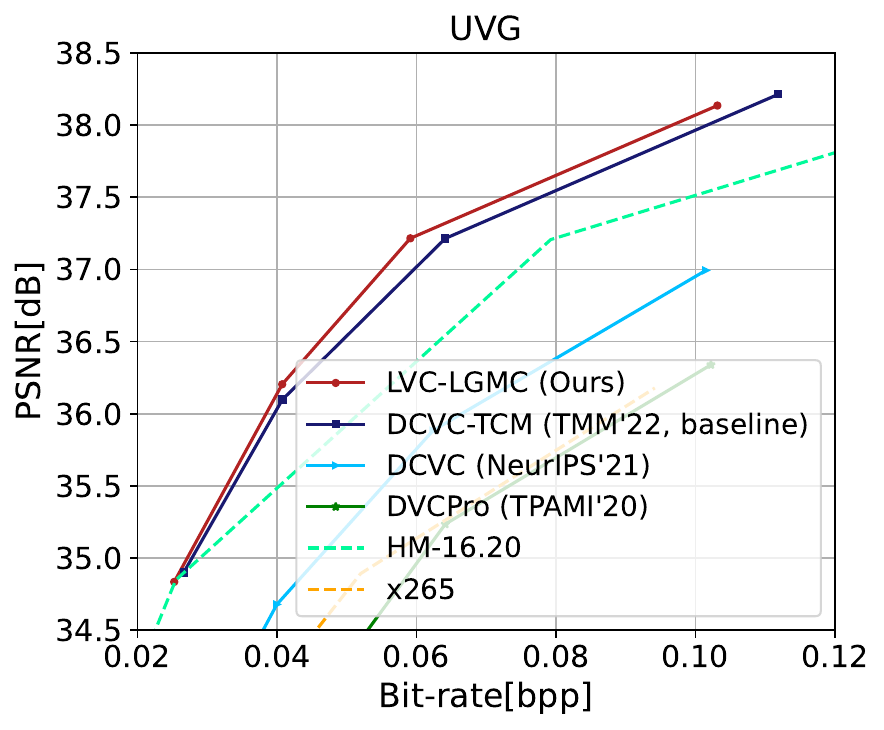}}
    \subfloat{
        \includegraphics[scale=0.35]{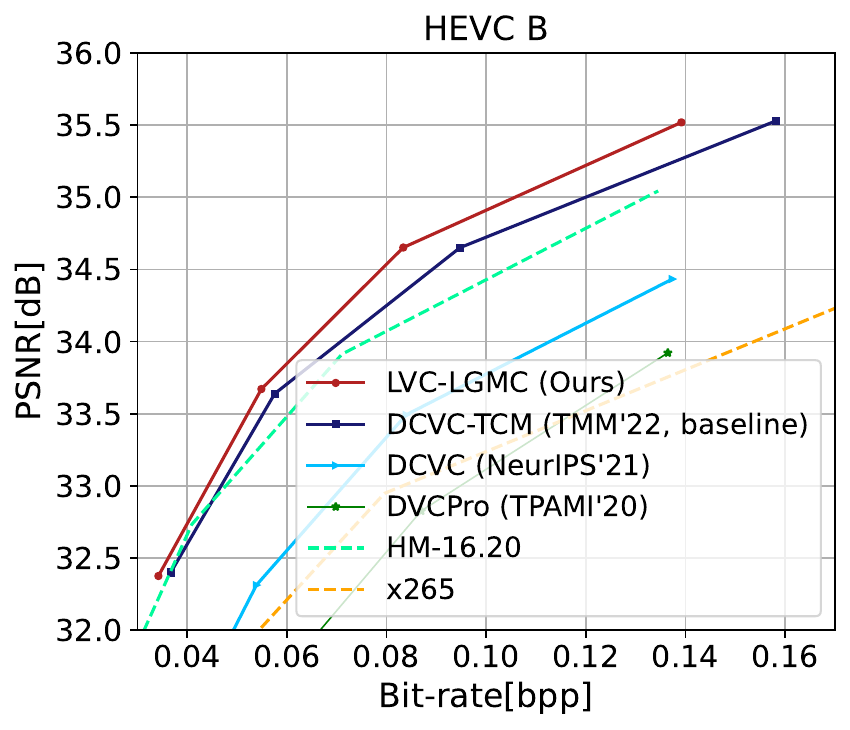}}\\
    \subfloat{
        \includegraphics[scale=0.35]{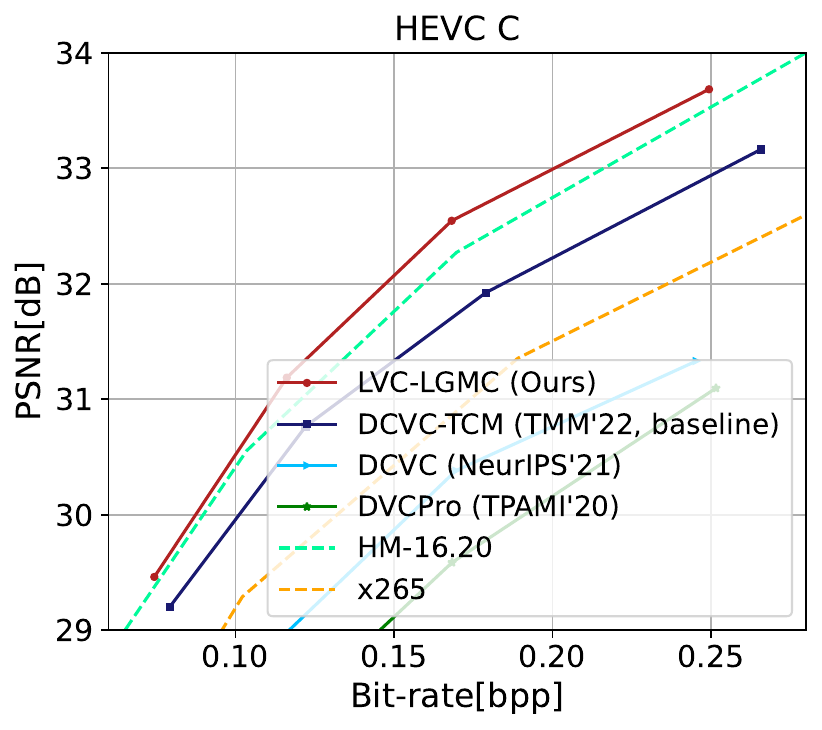}}
    \subfloat{
        \includegraphics[scale=0.35]{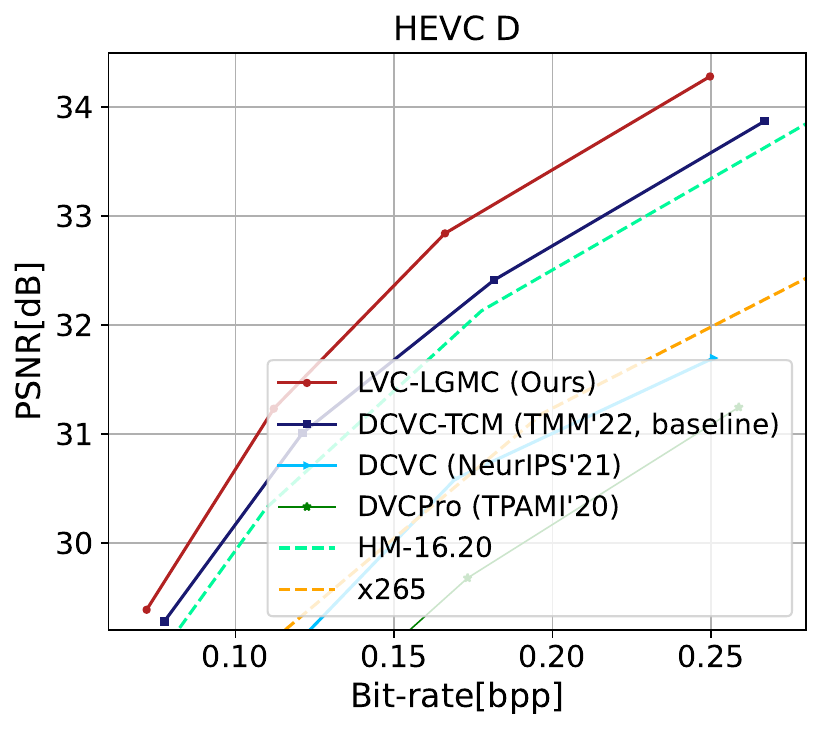}}
\caption{Illustration of rate-distortion performance of the proposed LVC-LGMC, DCVC-TCM~\cite{sheng2022temporal}, DCVC~\cite{li2021deep}, DVCPro~\cite{lu2020end}, HM-16.20 and x265 codec. 
The distortion is PSNR.
Please zoom in for better view.
    }
    \label{fig:rd_psnr}
\end{figure*}
The $\textrm{softmax}\left(\boldsymbol{y}_t^2 \left(\hat{\boldsymbol{f}}_t^2\right)^\top\right) \in \mathbb{R}^{\frac{L}{16}\times \frac{L}{16}} \geq 0$, and it can 
be treated as the similarity metric. It computes the similarity between a symbol
and all other symbols, such that the similarity can be captured.
The attention and $\boldsymbol{y}_t^2$ are then concatenated to 
conduct conditional coding on the basis of global
dependency.\par
However, the computational complexity of Eqn.~\ref{eq:vanilla} is $O(L^2)$.
The quadratic complexity makes it hard to employ the \textit{vanilla} approach for high-resolution video coding.
The quadratic is caused by the softmax operation, which specifies the order of matrix multiplication.
To solve the quadratic complexity, we employ efficient attention~\cite{shen2021efficient, jiang2023mlic++}, which
employs the softmax operation on $\boldsymbol{y}_t^2$ in row and the softmax operation on $\hat{\boldsymbol{f}}_t^2$ in column,
\begin{equation}
    \hat{\boldsymbol{g}}_{t}^2 = \underbrace{\textrm{softmax}\left(\boldsymbol{y}_t^2\right) \textrm{softmax}\left(\hat{\boldsymbol{f}}_t^2\right)^\top}_{\textrm{non-negative}}\hat{\boldsymbol{f}}_t^2.
\end{equation}
Since $\textrm{softmax}\left(\boldsymbol{y}_t^2\right) \textrm{softmax}\left(\hat{\boldsymbol{f}}_t^2\right)^\top \geq 0$, 
which makes is can be used as a similarity metric. Larger values denotes higher similarity.
In practice, $\textrm{softmax}\left(\hat{\boldsymbol{f}}_t^2\right)^\top\hat{\boldsymbol{f}}_t^2$ is computed first, resulting in the $O(C^2L)$ complexity.\par
During decompression, the attention-based global compensation is also conducted.
When conducting global motion Compensation, the overall process is 
\begin{equation}
    \hat{\boldsymbol{x}}_{t} = D_C\left(\left\lfloor E_C(\boldsymbol{x}_t|\hat{\boldsymbol{g}}_{t}^0,\hat{\boldsymbol{g}}_{t}^1,\hat{\boldsymbol{g}}_{t}^2)\right\rceil|\hat{\boldsymbol{g}}_{t}^0,\hat{\boldsymbol{g}}_{t}^1,\hat{\boldsymbol{g}}_{t}^2\right).
\end{equation}
\begin{figure*}[t]
    \centering
    \subfloat{
        \includegraphics[scale=0.35]{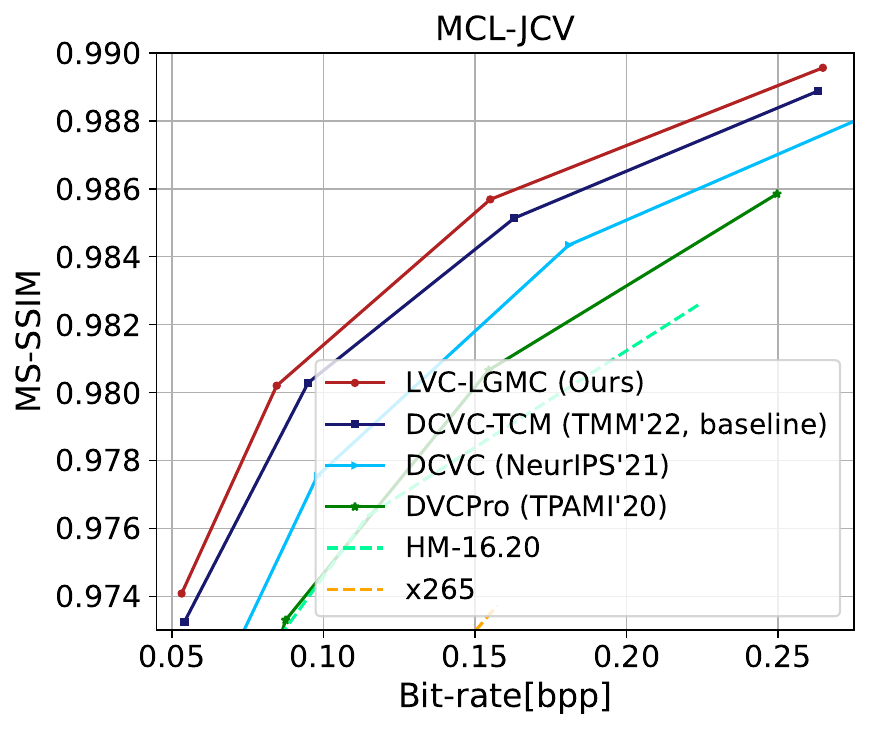}}
    \subfloat{
        \includegraphics[scale=0.35]{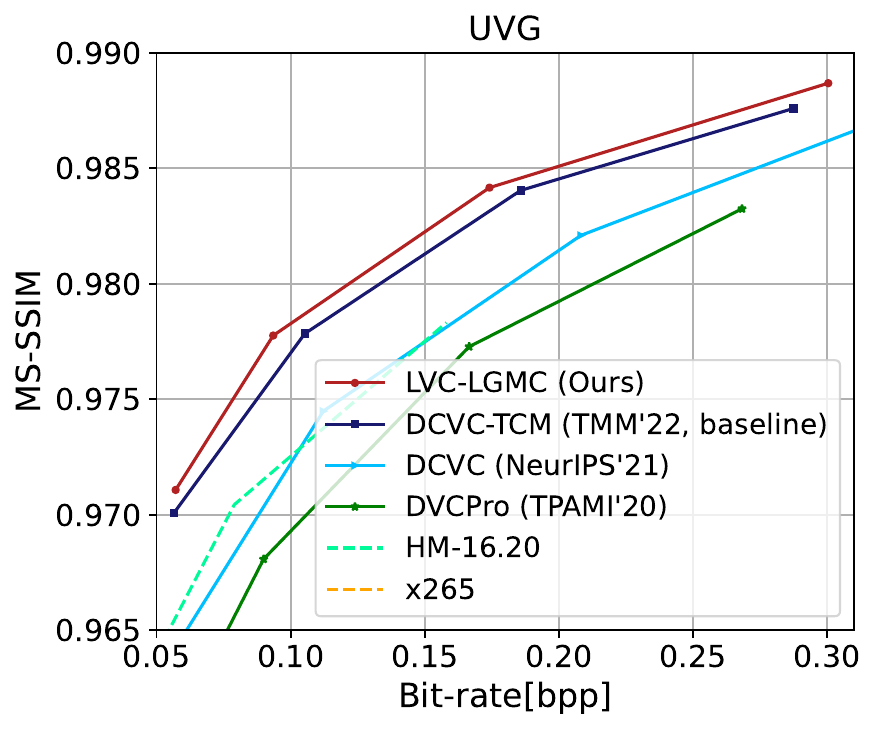}}
    \subfloat{
        \includegraphics[scale=0.35]{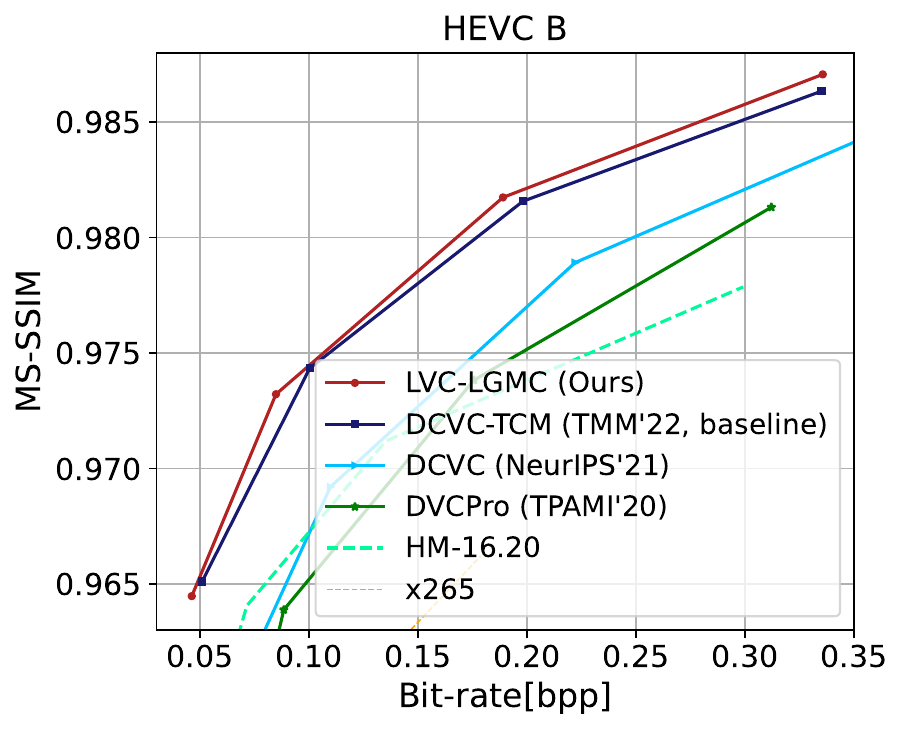}}\\
    \subfloat{
        \includegraphics[scale=0.35]{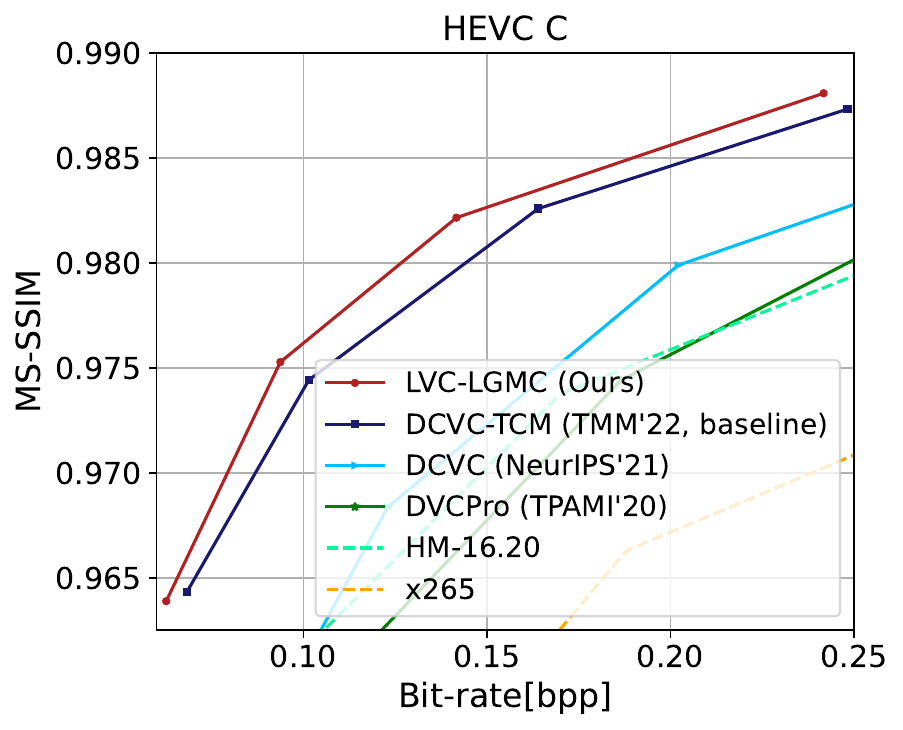}}
    \subfloat{
        \includegraphics[scale=0.35]{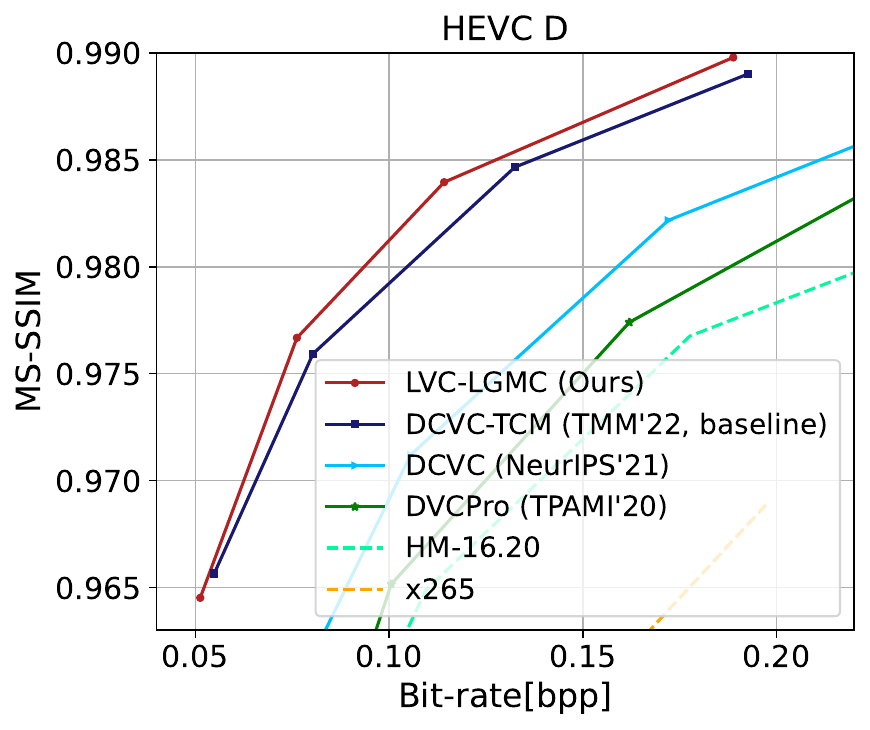}}
\caption{Illustration of rate-distortion performance of the proposed LVC-LGMC, DCVC-TCM~\cite{sheng2022temporal}, DCVC~\cite{li2021deep}, DVCPro~\cite{lu2020end}, HM-16.20 and x265 codec. 
The distortion metric is MS-SSIM~\cite{wang2003multiscale}.
Please zoom in for better view.
    }
    \label{fig:rd_ssim}
\end{figure*}
\begin{table*}[t]
    \footnotesize
    \centering
    \begin{tabular}{@{}cccccccccccc@{}}
    \toprule
    \multicolumn{1}{c|}{\multirow{2}{*}{Methods}}                            & \multicolumn{2}{c}{MCL-JCV~\cite{wang2016mcl}}       & \multicolumn{2}{c}{UVG~\cite{mercat2020uvg}}   & \multicolumn{2}{c}{HEVC B} & \multicolumn{2}{c}{HEVC C} & \multicolumn{2}{c}{HEVC D}   \\
    \multicolumn{1}{c|}{}                                                     & \multicolumn{1}{c}{PSNR} & \multicolumn{1}{c}{MS-SSIM} & \multicolumn{1}{c}{PSNR}& \multicolumn{1}{c}{MS-SSIM} & \multicolumn{1}{c}{PSNR}& \multicolumn{1}{c}{MS-SSIM}& \multicolumn{1}{c}{PSNR}& \multicolumn{1}{c}{MS-SSIM} & \multicolumn{1}{c}{PSNR}& \multicolumn{1}{c}{MS-SSIM} \\ \midrule
    \multicolumn{1}{c|}{HM-16.20}                     & 0.0      & 0.0       & 0.0       & 0.0 & 0.0 &  0.0 & 0.0  & 0.0 & 0.0 &  0.0 \\\midrule
    \multicolumn{1}{c|}{DVCPro~\cite{lu2020end} (TPAMI'20)}                     & 99.3      & 7.8       & 137.7       & 36.2 &123.7 &  23.5& 124.0  & 17.0 & 93.6 &  -7.8 \\\midrule
     \multicolumn{1}{c|}{RLVC~\cite{yang2020learning} (CVPR'20)}                     & 124.8      & 34.5       & 140.1       & 49.4 & 122.6 &  28.3 & 118.9  & 30.0 & 81.2 &  0.2 \\\midrule
    \multicolumn{1}{c|}{MLVC~\cite{lin2020mlvc} (CVPR'20)}                     & 66.8      & 50.3       & 66.5       & 64.7 &61.4 &  50.2& 124.1  & 53.1 & 96.1 &  40.4 \\\midrule
    \multicolumn{1}{c|}{DCVC~\cite{li2021deep} (NeurIPS'21)}                     & 42.8      & -16.3       & 67.3       & 9.2 &56.0 &  0.9& 76.9  & -8.9 & 52.8 &  -24.2 \\\midrule
    \multicolumn{1}{c|}{CANF-VC~\cite{ho2022canf} (ECCV'22)}                     & 8.5      & -21.9       & 6.2       & -4.4 &9.0 &  -6.8& 21.0  & -9.4 & 12.5 &  -18.1 \\\midrule
    \multicolumn{1}{c|}{DCVC-TCM~\cite{sheng2022temporal} (TMM'22)}                     & -3.2      & -38.3       & -9.0       & -25.5 &-5.3 &  -40.8& 15.1  & -42.4 & -5.4 &  -52.6 \\\midrule
    \multicolumn{1}{c|}{LVC-LGMC (Ours)} & \textbf{-13.0}      & \textbf{-44.3}       & \textbf{-13.2}              &\textbf{-32.4}      &\textbf{-13.1}  &\textbf{-44.8}  &\textbf{-5.3}  &\textbf{-48.4}  &\textbf{-19.1}   &\textbf{-56.3} \\\bottomrule
    \end{tabular}
    \caption{BD-rate (\%) for PSNR and MS-SSIM~\cite{wang2003multiscale}. 
    }
    \label{tab:rd}
  \end{table*}
\subsection{Joint Local and Global Motion Compensation for Learned Video Compression}
Our proposed joint local and global motion compensation (LGMC)
is illustrated in Fig.~\ref{fig:lmc} and Fig.~\ref{fig:lmc_dec}. Flow is employed to warp to obtain
local contexts $\hat{\boldsymbol{l}}_{t}^0,\hat{\boldsymbol{l}}_{t}^1,\hat{\boldsymbol{l}}_{t}^2$ and 
attention is used to obtain global contexts $\hat{\boldsymbol{g}}_{t}^0,\hat{\boldsymbol{g}}_{t}^1,\hat{\boldsymbol{g}}_{t}^2$.
We concatenate the local context, global context with current frames or 
middle feature. In this manner, local and global contexts can be utilized for 
conditional coding. The overall process can be described as follows, 
\begin{equation}
    \begin{aligned}
        \hat{\boldsymbol{y}}_t &= \lfloor E_C(\boldsymbol{x}_t|\hat{\boldsymbol{l}}_{t}^0,\hat{\boldsymbol{l}}_{t}^1,\hat{\boldsymbol{l}}_{t}^2,\hat{\boldsymbol{g}}_{t}^0,\hat{\boldsymbol{g}}_{t}^1,\hat{\boldsymbol{g}}_{t}^2)\rceil,\\
        \hat{\boldsymbol{x}}_{t} &= D_C\left(\hat{\boldsymbol{y}}_t
         |\hat{\boldsymbol{l}}_{t}^0,\hat{\boldsymbol{l}}_{t}^1,\hat{\boldsymbol{l}}_{t}^2,\hat{\boldsymbol{g}}_{t}^0,\hat{\boldsymbol{g}}_{t}^1,\hat{\boldsymbol{g}}_{t}^2\right),
    \end{aligned}
    \end{equation}
where $\hat{\boldsymbol{y}}_t$ is the latent representation of $\boldsymbol{x}_t$.
\section{Experiments}
\label{sec:exp}
\subsection{Experimental Setup}
\label{sec:exp:exp_setup}
We train the LVC-LGMC on Vimeo-90k dataset~\cite{xue2019video}. 
Frames are randomly cropped into $256\times 256$ patches.
The proposed models are optimized with the rate-distortion loss as follows,
\begin{equation}
\begin{aligned}
    \mathcal{L} &= \mathcal{R} + \lambda \times \mathcal{D},
\end{aligned}
\end{equation}
where $\mathcal{D}$ denotes the distortion metric Mean Square Error (MSE) or MS-SSIM~\cite{wang2003multiscale}. To cover wide range of coding rates, the $\lambda$ is set as $\{256, 512, 1024, 2048\}$ for MSE and $\{256, 512, 1024, 2048\} \times \frac{1}{50}$ for MS-SSIM.
Following DCVC-TCM~\cite{sheng2022temporal}, 
we adopt the multi-stage
training strategy with the AdamW optimizer, and the batch size is set as $4$. 
\subsection{Rate Distortion Performance}
\label{sec:exp:per}
To demonstrate the effectiveness of the proposed method, the rate-distortion performance and model complexity is evaluated.
We compare our LVC-LGMC with existing video coding schemes, including DCVC-TCM~\cite{sheng2022temporal}, CANF-VC~\cite{ho2022canf},
DCVC~\cite{li2021deep}, MLVC~\cite{lin2020mlvc}, RLVC~\cite{yang2020learning}, DVCPro~\cite{lu2020end}, HM-16.20, x265. The UVG~\cite{mercat2020uvg},
MCL-JCV~\cite{wang2016mcl}, and HEVC test sequences (class B, class C and class D) are all involved for performance evaluation.
In particular, for learned coding schemes, we employ the official models for testing.
First 96 frames in each sequences are involved in testing, and the intra period is set to 32. The height and width of frames are padded to the multiples of 64 to facilitate testing.
Fig.~\ref{fig:rd_psnr}, Fig.~\ref{fig:rd_ssim} and Table~\ref{tab:rd} present the rate-distortion performance.
The proposed global and local motion compensation model LVC-LGMC
outperforms the baseline DCVC-TCM model~\cite{sheng2022temporal} on all datasets,
indicating the effectiveness of employing global and local motion compensation for motion estimation in learning-based video coding.
Moreover, the proposed 
LVC-LGMC shows superior rate-distortion performance when compared with the CANF-VC, DCVC, MLVC, RLVC, DVCPro, HM-16.20, and x265.
When compared with baseline model DCVC-TCM,
the proposed LVC-LGMC achieves significant bit-rate savings. More specifically, the BD-rate gains on MCL-JCV are 10.09\% in terms of PSNR and 9.74\% BD-rate savings are obtained in terms of MS-SSIM. Moreover, it is interesting to see that we could constant achieve the rate-distortion performance improvement on different resolutions and contents, indicating the effectiveness of the proposed method.  
We compare the LVC-LGMC with the baseline model DCVC-TCM
regarding the parameter numbers, and coding complexity. The test environment is Tesla A100 GPU. The parameter numbers of the proposed 
LVC-LGMC and DCVC-TCM are 14.09M and 10.71M, respectively. 
The coding complexity increament of the LVC-LGMC is marginal, which increases the decoding time by 20\% on 1080p sequences.
The complexity increases of the LVC-LGMC is tolerable as the rate-distortion performance improvements are significant. 
  \begin{figure}[t]
    \centering
    \subfloat{
      \includegraphics[scale=0.2]{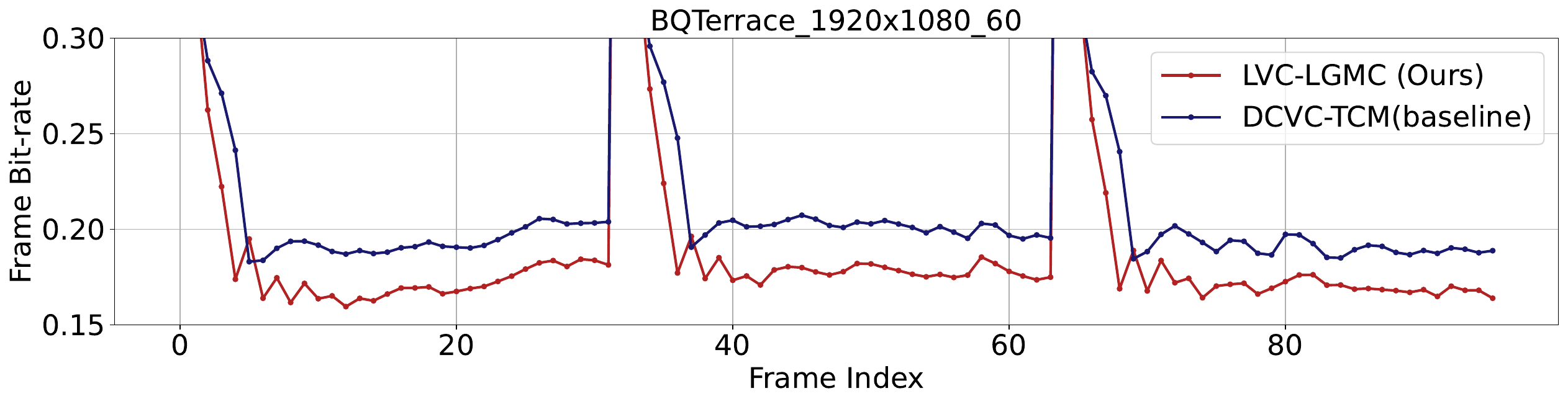}}\\
    \subfloat{
      \includegraphics[scale=0.2]{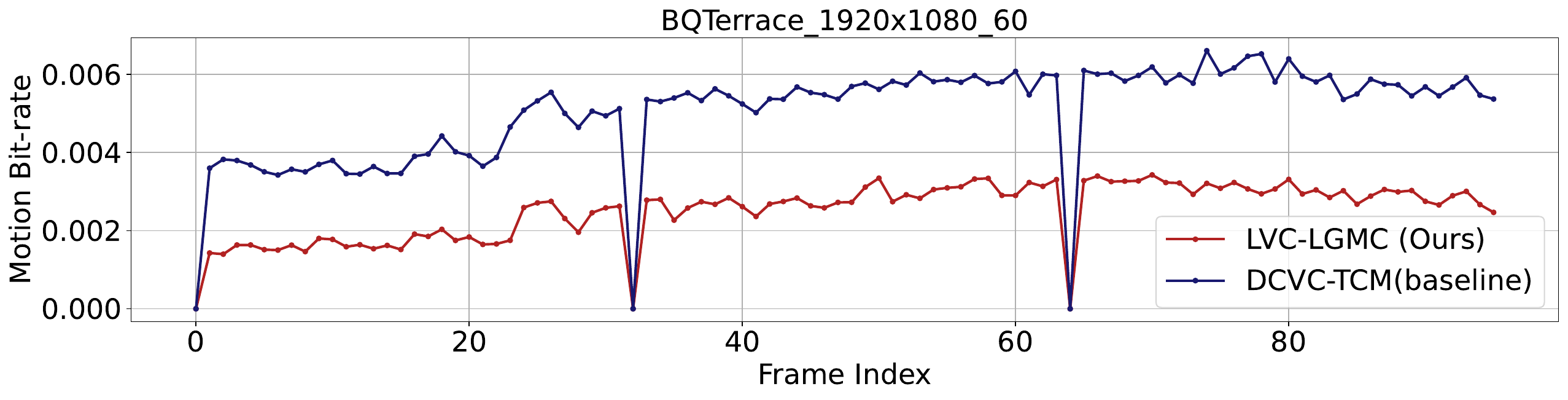}}
    \caption{Bit allocation of LVC-LGMC and baseline model DCVC-TCM~\cite{sheng2022temporal}.}
    \label{fig:bpp_change}
\end{figure}
\begin{figure}
    \centering
    \includegraphics[width=\linewidth]
    {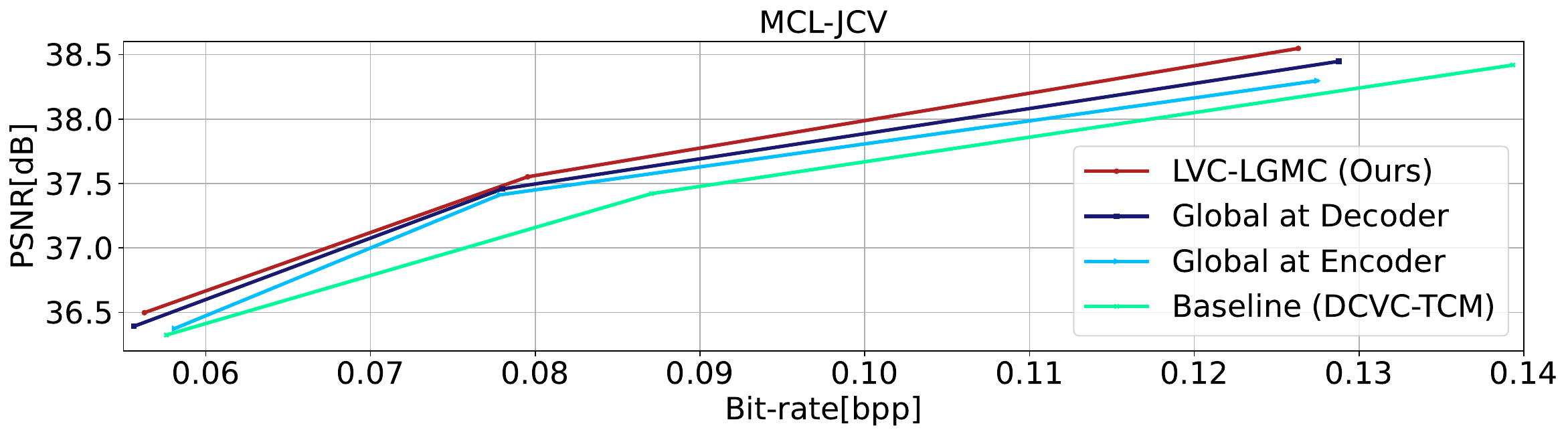}
    \caption{Ablation Studies on MCL-JCV dataset.}
    \label{fig:ablation}
\end{figure}
\subsection{Analyses and Discussions}
\label{sec:analysis}
\subsubsection{Bit Allocation}
We analyze the proposed mixed attention-flow based motion estimation and 
compensation via bit-allocation. We take sequence BQTerrace from 
HEVC B as an example, where the resolution is 1920$\times$1080 and the frame rate is 60. The average PSNR of LVC-LGMC is 33.878 dB and the average PSNR of DCVC-TCM is 33.871 dB.
The associated bit-rate and motion bit-rate of individual frame are presented in Fig.~\ref{fig:bpp_change}.
The proposed LVC-LGMC consumes lower coding bits in motion representation, and the overall bit consumption is lower than the DCVC-TCM. 
It is worthy of noting that the proposed global module is bit-free for motion representation.
Long-range similarities can be well estimated with the global module, such that fewer bits are required for representing the flow map. Consequently, the enhanced prediction leads to the overall bit reduction of the frame-level information coding. 
\subsubsection{Ablation Studies}
To comprehensively evaluate the effectiveness of the proposed attention-based global 
context for motion compensation, we remove the global module at the encoder-side
and remove the global module at the decoder-side.
Removing global at the encoder-side is quite similar with distributed source coding. 
The results are presented in Fig.~\ref{fig:ablation}.
Global at encoder and global at decoder both lead to performance degradation, but 
performs better than baseline DCVC-TCM~\cite{sheng2022temporal}.
\section{Conclusion}\label{sec:conclu}
In this paper, we propose joint global and local motion compensation (LGMC) for learned video coding.
We propose to employ the cross attention to capture global contexts for global motion compensation.
To reduce the complexity, we divide the 
softmax operation in vanilla attention into two independent 
softmax operations, leading to linear complexity.
We employ existing flow-based motion comprensation for local contexts.
To evaluate our proposed module, we incorporate it with DCVC-TCM
and we get video compression model LVC-LGMC. Extensive experiments
demonstrate that our LVC-LGMC have significant improvements over corresponding baseline DCVC-TCM.
Our methods are plug-and-play and can be employed in other conditional coding based models~\cite{li2021deep, li2022hybrid, li2023neural} for
further performance improvements.

\bibliographystyle{IEEEbib}
\bibliography{refs}
\end{document}